# On Tsunami Waves induced by Atmospheric Pressure Shock Waves after the 2022 Hunga Tonga-Hunga Ha'apai Volcano Eruption

Zhiyuan Ren[1], Pablo Higuera[1,2], Philip Li-Fan Liu[1,3,4,5]

[1]Department of Civil and Environmental Engineering, National University of Singapore, Singapore

[2]Department of Civil and Environmental Engineering, The University of Auckland, New Zealand

[3]School of Civil and Environmental Engineering, Cornell University, Ithaca, USA

[4]Institute of Hydrological and Oceanic Sciences, National Central University, Taiwan

[5]Department of Hydraulic and Ocean Engineering, National Cheng Kung University, Taiwan

Corresponding author: phigueracoastal@gmail.com

**Key Points**

1. Using the measured data, an N-wave shaped atmospheric pressure model is constructed for the 2022 Tonga event.

2. The atmospheric pressures generate both locked waves, propagating at the speed of pressure, and trailing free waves with a long wave speed.

3. The trailing free tsunami waves can be as significant as the leading locked tsunami waves, which are amplified by the Proudman resonance.

Confidential manuscript submitted to *Journal of Geophysical Research: Oceans*

**Abstract**

Employing a linear shallow water equation (LSWE) model in the spherical coordinates, this paper investigates the tsunami waves generated by the atmospheric pressure shock waves due to the explosion of the submarine volcano Hunga Tonga–Hunga Ha'apai on January 15, 2022. Using the selected 59 atmospheric pressure records in the Pacific Ocean, an empirical atmospheric pressure model is first constructed. Applying the atmospheric pressure model and realistic bathymetric data in the LSWE model, tsunami generation and propagation are simulated in the Pacific Ocean. The numerical results show clearly the co-existence of the leading locked waves, propagating with the speed of the atmospheric pressure waves (~ 1,100 km/hr), and the trailing free waves, propagating with long gravity ocean wave celerity (~ 750 km/hr). During the event, tsunamis were reported by 41 DART buoys in the Pacific Ocean, which require corrections because of the occurrence of atmospheric pressure waves. The numerically simulated tsunami arrival time and the amplitudes of the wave crest and trough of the leading locked waves compare reasonably well with the corrected DART measurements. The comparisons for the trailing waves are less satisfactory, since free waves could also have been generated by other tsunami generation mechanisms, which have not been considered in the present model, and by the scattering of locked waves over changing bathymetry. In this regard, the numerical results show clearly that the deep Tonga trench (~ 10 km) amplifies the trailing waves in the Southeast part of the Pacific Ocean via the Proudman resonance condition.

**Key words**

2022 Tonga Tsunami; Atmospheric Pressure function; Meteotsunami; Numerical Simulation

**Plain Language Summary**

On January 15, 2022, the eruption of the submarine volcano Hunga Tonga–Hunga Ha'apai, the largest volcano eruption in recent centuries, produced worldwide atmospheric pressure fluctuations and tsunami waves. In this paper we produced a simple model for atmospheric pressure in the Pacific Ocean based on selected pressure recordings. Based on this pressure model and realistic bathymetric data, we simulate the tsunami generation and propagation in the Pacific Ocean. We found that the explosion shock wave induced by the volcanic eruption is the main mechanism for tsunami generation. Two separate tsunami waves, travelling at different speeds, can be distinguished. Additional tsunami waves are also generated when the pressure wave travels over steep deep ocean features such as the Tonga Trench, leading to significantly larger waves in the Southeast part of the Pacific Ocean.

## 1 Introduction

On January 15, 2022 at approximately 04:15 (UTC), the submarine volcano Hunga Tonga–Hunga Ha'apai, located on the Tongan archipelago, exploded violently. The eruption produced a plume that elevated 58 km into the troposphere and an atmospheric pressure shock wave, propagating at a speed close to the sound speed, ~306-320 m/s (Amores et al., 2022; Burt, 2022; Matoza et al., 2022; Yuen et al., 2022). These atmospheric pressure waves were captured at weather stations around the world and were found circling the Earth at least 5 times over several days following the explosion (Amores et al., 2022). In addition, sea level oscillations were also observed all around the world. In the Pacific Ocean, DART buoys recorded tsunami waves with amplitudes ranging from a few centimeters to twenty centimeters (https://nctr.pmel.noaa.gov/Dart/). The coastal (tidal) gauges also recorded tsunami waves (http://www.ioc-sealevelmonitoring.org/). In the region far away from Tonga, the tsunami amplitudes were found to exceed 1 m (e.g., Chile, Japan, and USA). Tsunami waves were also captured in the Atlantic Ocean and Mediterranean Sea, even though they are not directly connected to the Pacific Ocean. Despite the complexity of the problem, theoretical studies (e.g., Liu and Higuera, 2022) have illustrated the basic physical mechanisms by which under the moving atmospheric pressure forcing the leading (locked) and trailing (free) waves can be simultaneously produced. While the locked waves propagate with the speed of the atmospheric pressure waves (~ 1,100 km/hr), the trailing free waves propagate with long gravity ocean wave (~ 750 km/hr) (e.g., Omira et al. 2022). Moreover, additional free waves are generated when the locked waves propagate over large bathymetric variations, such as trenches and seamounts (Vennell 2007). Lynett et al. (2022) argued that the trailing waves may also be associated with other generation mechanisms directly linked with the volcanic explosion, such as the collapse of the caldera.

Historically, there have been several tsunami events associated with volcanic eruptions. The most recent event is the 2018 Anak Krakatau eruption. During the eruption, the cone collapsed, leading to a landslide that generated tsunami waves (Ren et al., 2020; Grilli et al., 2021). The runup in the 2018 event exceeded 10 m along the coasts of the Sunda Strait, but the impact of the tsunami was local. No atmospheric pressure wave was recorded. On the other hand, the 1883 Krakatau volcano explosion generated significant atmospheric pressure waves, which were recorded in Europe and the United States, lasting for nine days (Choi et al., 2002). The tsunami waves were complex in the near field, since there were at least three possible generation mechanisms: caldera collapse, submarine explosion, and pyroclastic flow (Maeno and Imamura, 2011). In the far field, the correlations between the measured atmospheric pressure waves and tsunami waves suggested that the atmospheric pressure waves were the possible origin of the tsunami waves (Francis, 1985). Omira et al. (2022) has provided information on additional historical volcanic explosions that are linked to similar tsunami generation mechanisms.

For the 2022 Tonga event, abundant atmospheric pressure data and tsunami records are available in the Pacific Ocean region. The objectives of this paper are to correlate these two sets of data and understand the main tsunami wave features generated by the moving atmospheric pressure. To achieve these goals, we first construct an empirical model to describe the characteristics of the atmospheric pressure in time and space using the measured atmospheric pressure data in the Pacific Ocean after Tonga's volcanic explosion. This empirical atmospheric model is then used as the forcing function in a linear shallow water equation model to generate tsunami waves. The numerical results are compared with measurements, including the DART buoy data, which require a correction because the appearance of atmospheric pressure (Liu and Higuera, 2022). The wavelet analysis is applied to the corrected DART data and the numerical results to reveal the time evolution of the dominant wave components of the free surface elevation at selected DART stations. The dispersion effect on the tsunami propagation induced by the moving atmospheric pressure shock wave is discussed. The results of Lynett et al. (2022), which used a similar method, also are compared with our results.

This paper is structured as follows. Section 2.1 describes the procedure of constructing the empirical atmospheric pressure model. Section 2.2 explains briefly the reason why the DART buoy data needs to be corrected when the atmospheric pressure wave appears. Section 2.3 outlines the LSWE model setup. Section 3 is devoted to the analysis of the numerical modelling results, compared against the field observations. In addition, the role that dispersion plays in the pressure-driven tsunami propagation and the main differences with respect to the atmospheric model presented in Lynett et al. (2022) are also studied in this section. Finally, conclusions are drawn in Section 4.

## 2 Data and Methods

2.1 Atmospheric Pressure Observations and Empirical Model

After Tonga's volcanic eruption, thousands of Meteorological stations around the world captured the atmospheric pressure shock wave. These observations could be seriously affected by the location of the instrument and its surrounding topography. For example, several stations located in the cities of New Zealand present significantly smaller pressure peaks than those located near the coast. Moreover, the atmospheric pressure in Santiago de Chile presents overly high values, which could be caused by the Andes Mountains (reaching 6,500 m above sea level) partially reflecting the atmospheric pressure wave. Therefore, to construct an empirical atmospheric pressure model in the Pacific Ocean, we select atmospheric pressure stations based on two principles: (1) The instrument must be located at the coast, and (2) they need to be far from sharp orographic features (e.g., mountains or valleys).

In total, we have selected 59 stations, as shown in Figure 1, over three regions. The near field region is defined as the area within 3,000 km from Tonga, and comprises 16 stations (Fiji Meteorological Service, Weather Underground, NIWA), providing atmospheric pressure measurements with a 5-10 min sampling frequency. The mid-field region is within 3,000-6,000 km, and the far-field region is farther than 6,000 km from Tonga. There are 43 additional stations within these two regions, including 10 stations from Japan Soratena, 15 stations from the Meteorological Directorate of Chile, 2 stations from Taiwan, and 16 stations from the National Climatic Data Center of USA, providing a 1 min sampling frequency. Higher sampling rates are more likely to capture the peak values of the atmospheric pressure waves, thus, providing a more accurate representation of the event.

Confidential manuscript submitted to *Journal of Geophysical Research: Oceans*

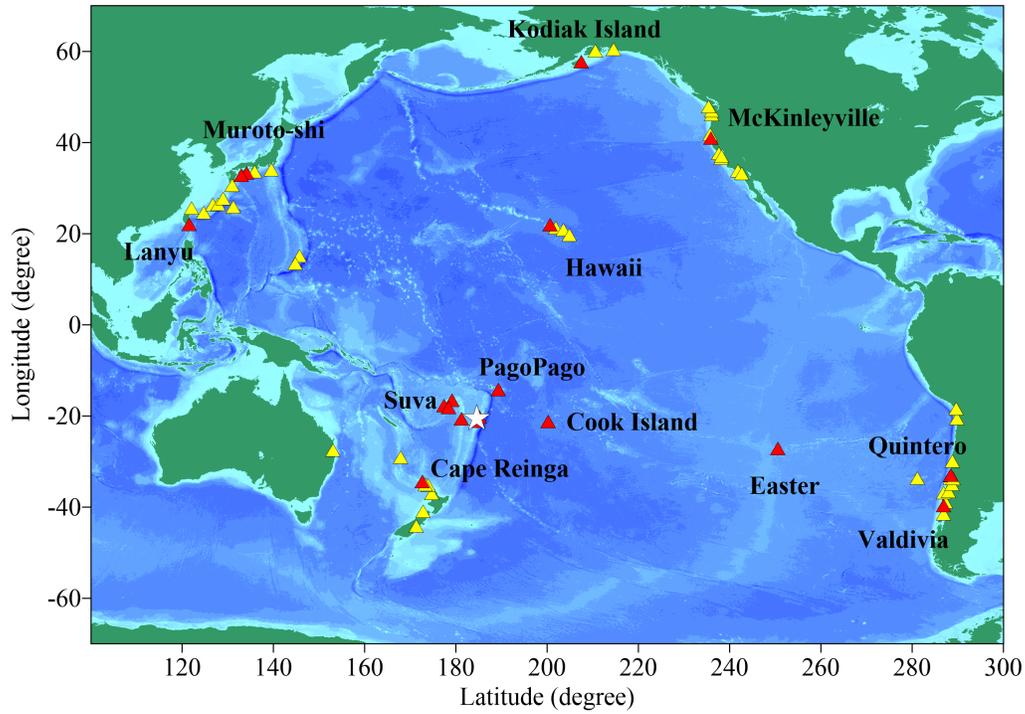

**Figure 1** Locations of the 59 atmospheric pressure stations, marked as triangles, around the Pacific Ocean being used in this paper. The 18 stations, highlighted in red, are those selected for comparison in Figure 5. The Hunga Tonga-Hunga Ha'apai volcano is denoted with a white star.

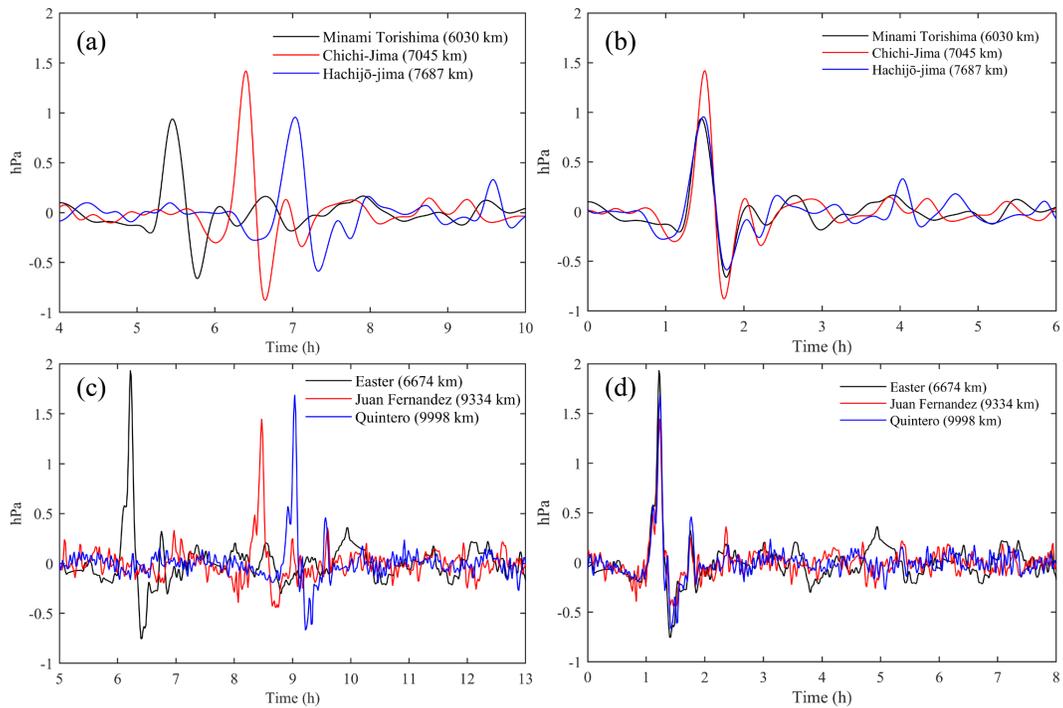

Figure 2 Similarity of atmospheric pressure data at different locations. The upper panel (a) plots the time histories of atmospheric pressure at three stations near Japan, and (b) shows the collapsed time histories when the arrival times are aligned . The lower panels (c) and (d) show the same features for three stations near Chile. Time in panels (b) and (d) is relative to the arrival time.

The atmospheric pressure data exhibits strong similarity in the far field. As shown in Figure 2, the time series of atmospheric pressure at three stations near Japan and Chile, present a distinctive N-wave shape when the arrival times are aligned, with the positive pressure at the crest being higher than the negative pressure at the trough. These stations are more than 6,000 km away from the source of explosion. The N-wave shape similarity is less obvious in the near field. Some of the near field data (e.g., Nomuka and Nukalofa, located less than 100 km from the source) do not show a clear crest, but are instead dominated by a deep trough, as shown in Figure 5 (black line). In the mid- and far-field the amplitudes of the atmospheric pressure wave crest and trough decay slowly with the distance to the source, as indicated in Figure 3. However, in the near-field, the pressure wave crest grows rapidly (see panel 3b). Moreover, the speed of the atmospheric pressure wave propagation at any atmospheric pressure sensor station can also be estimated. In this paper, the average speed, $\bar{c}(t)$, is calculated by dividing the distance from the source to the atmospheric pressure station by the travel time, which is the time when the atmospheric pressure signal first crosses the 0.001 hPa mark ahead of the arrival of the first pressure peak. The initial time of the event is set at 04:15 on 15th of January 2022. The estimated average wave celerity of the pressure wave as a function of the distance from the volcano is plotted in the left panel of Figure 4, showing that the atmospheric pressure speeds up quickly in the near field and reaches a constant value of ~319 m/s. The right panel of Figure 4 shows the same datapoints, but as a function of the time elapsed from the explosion, based on the arrival (i.e., travel) time definition.

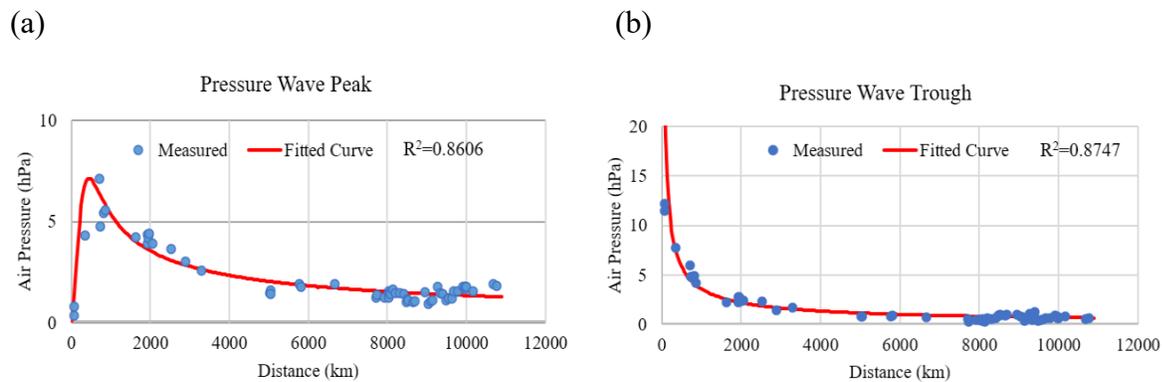

Figure 3 The variations of the crest and trough values of the atmospheric pressure with distance from the source of explosion.

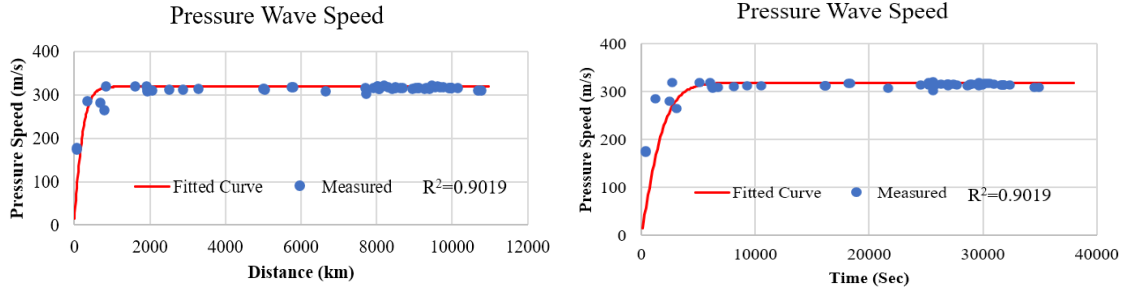

**Figure 4** The evolution of the average speed of the atmospheric pressure wave as a function of the distance from the source of explosion (left panel) and the time from the explosion (right panel).

Based on the available data, an empirical atmospheric pressure wave model in the spherical coordinate system is proposed as follows:

$$p_a(r, t) = A(X_1(t)) N(r, t), \qquad (1)$$

where

$$N(r, t) = \frac{3}{2}\sqrt{3} \operatorname{sech}^2[\Upsilon(r - X_1(t))]\tanh[\Upsilon(r - X_1(t))], \qquad (2)$$

$$X_1(t) = \int_0^t c(t)dt \approx \bar{c}(t)t, \qquad (3)$$

in which $p_a$ denotes the atmospheric pressure disturbance produced by the explosion, measured in [hPa] relative to normal conditions, at a given distance from the source ($r$, measured in meters) and time from the event ($t$, in seconds). Using the spherical coordinate system in terms of longitude and latitude ($\psi_1$, $\varphi_1$), the great-circle distance between the target location and the position of Hunga Tonga-Hunga Ha'apai Volcano ($\psi_0 = 175.39°W$, $\varphi_0 = 20.546°S$) is calculated as:

$$r = 2R \arcsin\left(\sqrt{\sin^2(\tfrac{\varphi_1-\varphi_0}{2}) + \cos(\varphi_1)\cos(\varphi_0)\sin^2(\tfrac{\psi_1-\psi_0}{2})}\right), \qquad (4)$$

in which, $R$ is the Earth radius ($6.371 \times 10^6$ m).

As indicated in Equation (1), the atmospheric pressure is modeled as the product of two functions. The shape function, $N(r, t)$ in Equation (2), prescribes the N-wave shape (see Tadepalli and Synolakis, 1994). The amplitude function, $A(X_1(t))$, provides the variations for the peak and trough pressure values, as a function of the distance $X_1(t)$. As defined in Equation (3), $X_1(t)$ can either be defined integrating the instantaneous wave celerity, $c(t)$, or as a simple product using the more convenient average speed $\bar{c}$, which has been previously defined. In the shape function, $\varUpsilon$ represents the wavenumber, defined as $\varUpsilon = 2\pi/L$, with $L$ being the wavelength (in meters) of the N-wave. According to the measured data at the selected stations the wavelengths range from 800 to 1,050 km (Amores et al., 2022; Lynett et al., 2022). For simplicity, in the present model, a constant average wavelength of 900 km is adopted. We have performed sensitivity analyses using larger and smaller wavelengths, which lead to insignificant variations and equally good fit in terms of the free surface elevation.

As illustrated in Figure 3, the crests and troughs of atmospheric pressure waves decay at different rates. Accordingly, the amplitude functions, $A(X_1(t))$, for the crest and the trough of the atmospheric pressure waves are determined separately by fitting curves through the measurements. These empirical functions can be expressed as:

$$A = \begin{cases} 340 \tanh^2\left(\frac{X_1(t)}{300{,}000}\right)\left(\frac{X_1(t)}{1{,}000}\right)^{-0.6} & if\ N(r,t) > 0, \text{for wave crest,} \\ 450 \left(\frac{X_1(t)}{1{,}000}\right)^{-0.7} & if\ N(r,t) < 0, \text{for wave trough.} \end{cases} \quad (5)$$

Note that the $R^2$ values for the fitted curves are 0.8606 and 0.8747 for wave crest and wave trough amplitudes, respectively. Equations (5) yield the pressure amplitudes in hPa, while the distance $X_1(t)$ is in meters. The propagation distance $X_1$ is obtained from Equation (3) as a simple product using the average speed. As shown in Figure 4, the average atmospheric wave celerity can be obtained by curve-fitting too and represented by

$$\bar{c}(t) = 319 \tanh(t/2200) \quad (6)$$

which yields a velocity in m/s when the time is in seconds.

To illustrate the accuracy of the empirical atmospheric pressure model, we have selected measured time series data at 18 stations, ranging from the near field to the far-field, for comparison as shown in Figure 5. The comparisons for the rest of stations are presented in Figure S1 in the supplementary materials. The modeled atmospheric pressure waves preserve the arrival time and the main wave shape very well. However, the model ignores the trailing oscillatory wave components. In the far field, larger differences in arrival time appear due to the inhomogeneity of the pressure wave travelling speed (produced by differences of atmospheric air temperature, see Amores et al., 2022), which is not accounted for in the present model.

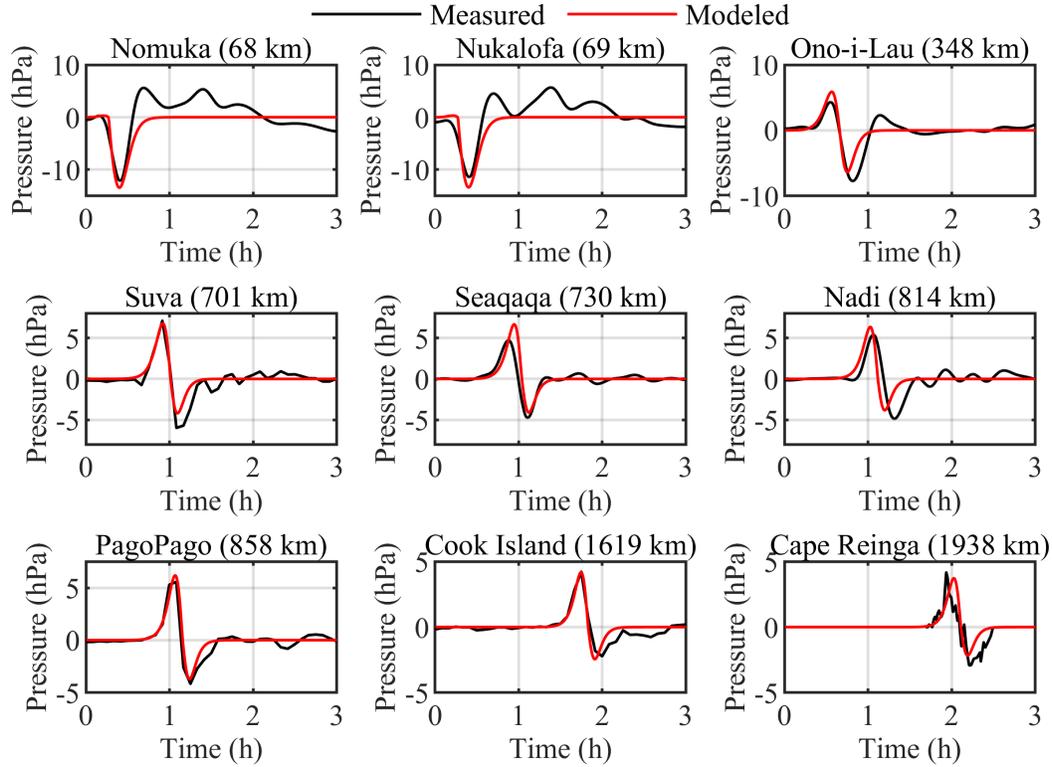

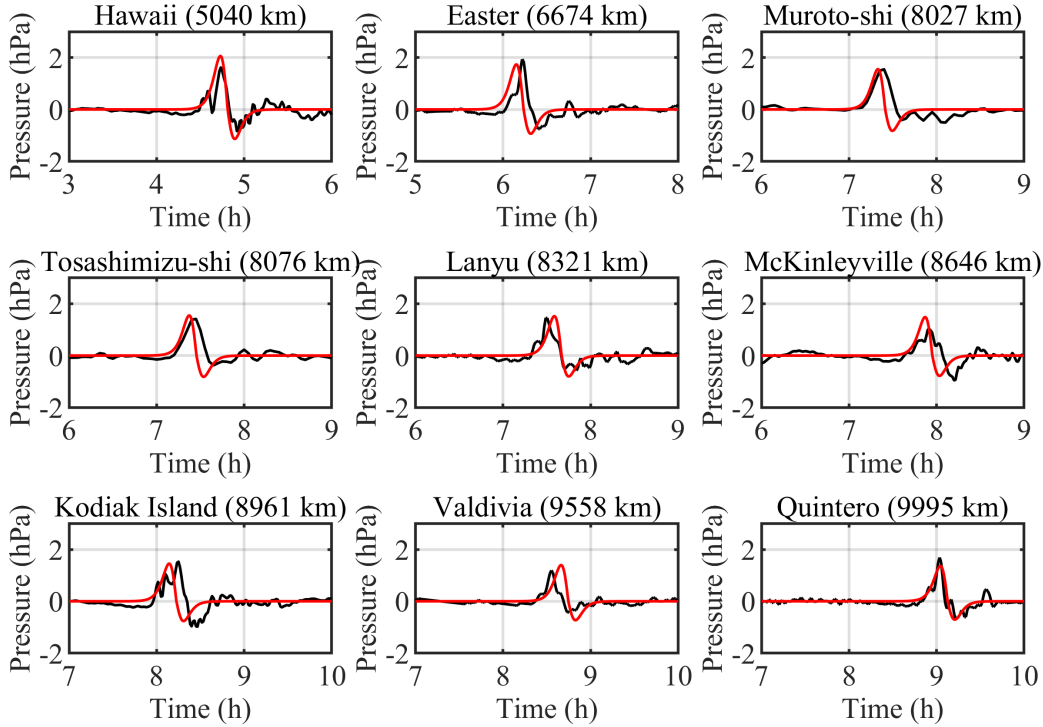

**Figure 5 Comparisons of time series of atmospheric pressure between measurements and model results at different stations**

2.2 Correction for DART Buoy data

In the Pacific Ocean, 41 DART buoy stations reported the sea level fluctuations. Normally, the DART buoy records the sea elevation data every 15 minutes. However, when the buoy detects a tsunami or abnormal sea level fluctuations, the sampling rate increases to 1 minute or 15 seconds (Meinig et al., 2005). Since the DART system measures the sea surface elevation using a bottom-mounted pressure recorder, its measurements ($\zeta$) include both atmospheric pressure fluctuations ($P_a$) and the sea level variation ($\eta$) (Rabinovich and Eblé, 2015),

$$\zeta = \eta + \frac{P_a}{\rho g} \qquad (7)$$

where $\rho$ is the density of seawater and $g$ is the gravitational acceleration. When tsunami waves are generated by an earthquake, the atmospheric pressure variations are negligible and the reported measurement ($\zeta$) is the same as the sea level variation ($\eta$). Moreover, DART stations do not record storm surge due to the "inverse barometer" sea level response to pressure. However, since the atmospheric pressure fluctuations induced by the Tonga volcano eruption are significant and the leading wave is "locked" in phase with the atmospheric pressure wave, the DART reported data must be corrected to isolate the actual tsunami waves. Based on the analysis of Liu and Higuera (2022), the formula to approximately correct the DART measurements can be expressed as

$$\eta = \frac{1}{F_r^2}\zeta, \qquad (8)$$

in which $F_r$ is the local Froude number at the DART buoy location, calculated as

$$F_r = \frac{c}{\sqrt{gd}}, \qquad (9)$$

where $c$ is the speed of the atmospheric pressure wave, which is fixed, and $d$ is the local water depth. It can be observed that the correction method is dependent on the local water depth, and results in a reduction of the DART reported data for $F_r > 1$, which is the case for the Tonga event. Moreover, this reduction (Equation 8) is linearly proportional to the water depth, thus, the formula will produce a relatively larger correction in shallower waters.

Figure 6 shows the location of the DART buoy stations across the Pacific Ocean (panel a) and corrected and un-corrected time series of free surface elevations at 4 selected DART buoy stations (panels b~e). These stations have been selected because the water depths cover a wide range: 1,807 m (DART52406), 3,273 m (DART32411), 4,230 m (DART46409) and 5,742 m (DART G), respectively. To correct the DART data, the atmospheric pressure at the DART buoy location is first estimated using the empirical atmospheric pressure model. The duration of the atmospheric pressure wave is extracted and used as a mask to correct the DART data only within the relevant window (shown in grey line), using Equation 8. The speeds of atmospheric pressure wave at each station are calculated, which in this case are almost a constant (~ 319 m/s), since these stations are farther than 1,000 km from Tonga. Therefore, the Froude number at these buoys ranges from 1.32 (DART52406) to 2.4 (DART G), depending on the local water depths. The blue line is the originally reported DART data and the black line denotes the corrected time series, i.e., the actual tsunami free surface elevation. It can be observed that, as mentioned before, the relative reduction is larger for shallower water depths. Panels b~e in Figure 6 are arranged in terms of increasing water depths. Therefore, the most significant reduction occurs for DART 52406, from 3.8 cm of positive wave amplitude reported to 0.7 cm after the correction.

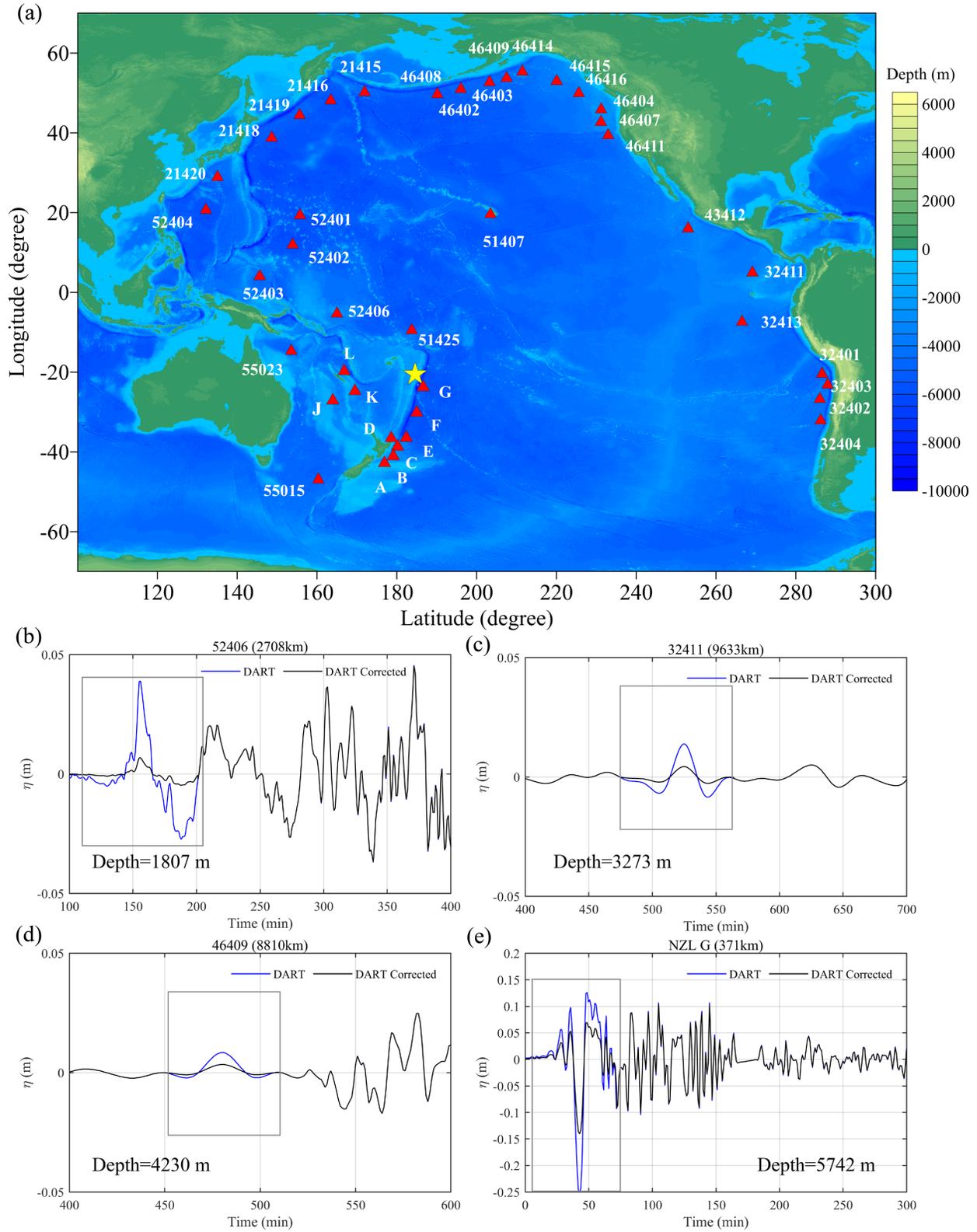

**Figure 6 Reported and corrected DART buoy data. The upper panel (a) shows the locations of the DART buoys in the Pacific Ocean. The lower panels show the time series of the surface elevation of four buoys, arranged in terms of increasing water depths, before (blue line) and after (black line) the correction has been applied within the grey window**.

2.3 Numerical Model

The numerical model FUNWAVE (Shi et al., 2012; Kirby et al., 2013) has been adopted for simulating tsunami generation and propagation using the linear shallow water equations (LSWE). In the spherical coordinates, the LSWE are written as,

$$\frac{\partial \eta}{\partial t} + \frac{1}{R\cos\varphi}\left\{\frac{\partial M}{\partial \psi} + \frac{\partial}{\partial \varphi}(\cos\varphi N)\right\} = 0 \qquad (10)$$

$$\frac{\partial M}{\partial t} + \frac{gd}{R\cos\varphi}\frac{\partial \eta}{\partial \psi} - fN = -\frac{d}{\rho R\cos\varphi}\frac{\partial p_a}{\partial \psi} \qquad (11)$$

$$\frac{\partial N}{\partial t} + \frac{gd}{R}\frac{\partial \eta}{\partial \varphi} + fM = -\frac{d}{\rho R}\frac{\partial p_a}{\partial \varphi} \qquad (12)$$

where $M$ and $N$ are the volume fluxes in the longitude ($\psi$) and latitude ($\varphi$) directions, respectively, and $f$ is the Coriolis force.

In the numerical model, the bathymetry data has been obtained from the ETOPO1 global relief model at 1 arcmin resolution. The computational domain spans from 100° E to 120° W, and from 70° S to 70° N (Figure 1), which covers the entire Pacific Ocean region. The grid size is 4 arcmin (~3,600 m), and the total grid cell count is 6.3 million. The time step is 6 s, which is small enough to satisfy the CFL condition. The propagation of the atmospheric pressure and the resulting tsunami waves are simulated for 20 hours from the volcanic explosion using 48 cores, and the simulations take 4.6 hours on a 2.4 GHz Xeon workstation. Additional grid refinement sensitivity tests have been carried out based on grid resolutions of 2 arcmin, 4 arcmin and 8 arcmin, respectively. For the 2 arcmin grid resolution, the sensitivity analysis could not be completed because significant numerical instabilities are observed. Nevertheless, the comparisons between the 4 and 8 arcmin grids indicated that they produced the same results in the near field and only slight differences were observed in the far-field. In the end, the 4 arcmin grid is adopted in this study.

# 3 Results and Discussions

## 3.1 Model/data comparisons

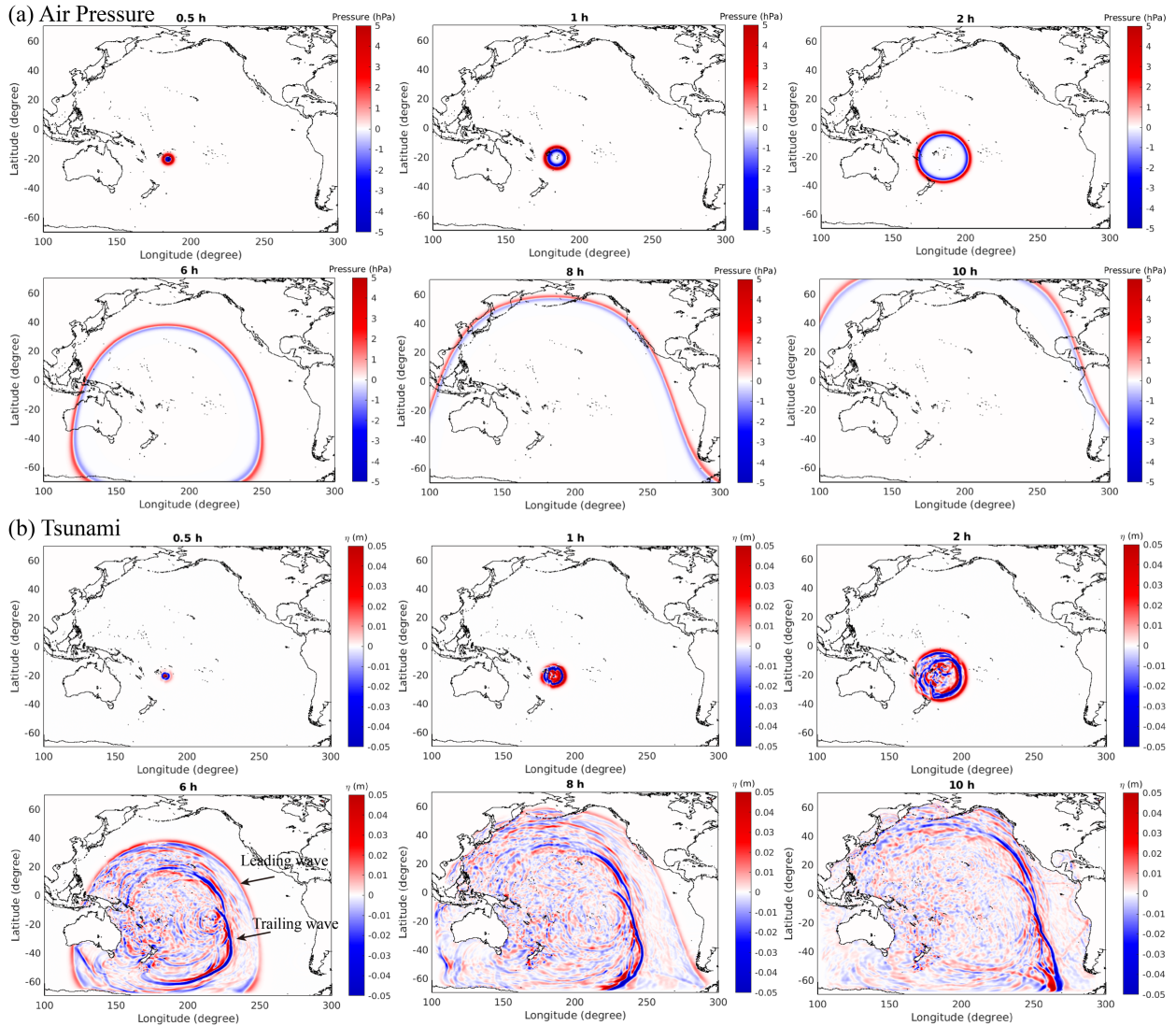

**Figure 7 Snapshots of (a) the atmospheric pressure waves and (b) the tsunami waves in the Pacific Ocean**

Figure 7 shows snapshots of modeled atmospheric pressure waves and the corresponding simulated tsunami waves in the Pacific Ocean, illustrating the similarities of the general pattern between two wave fields. The top panel (a) shows a sequence of the circular atmospheric pressure waves, being prescribed by the empirical N-wave model in Equation 1, which propagate across the Pacific Ocean at a speed of ~319 m/s. The amplitude of the atmospheric pressure decreases from the near-field to the mid-range-field, as specified by the model. Six hours after the explosion, the amplitude has reduced to approximately 1 hPa, and after 10 hours, the atmospheric pressure wave is already propagating away from the Pacific Ocean. The atmospheric pressure waves will eventually circle around the Earth and return to the Pacific Ocean several times. However, the present empirical model is only applicable to the initial stage covering the first pass over the Pacific Ocean.

The panel (b) in Figure 7 shows the corresponding snapshots of the simulated tsunami waves in the Pacific Ocean. The tsunami waves originate in the vicinity the volcano, where the amplitude is significantly larger due to the large atmospheric pressure in the near-field, which decays fast, as mentioned before. The snapshots of the tsunami at 1 h and 2 h show that the tsunami wave heights to the East of Tonga are larger than those to the West. This is due to Proudman resonance condition occurring over the Tonga Trench, where the Froude number $Fr = \frac{c}{\sqrt{gh}}$, is close to the critical condition. This point will be further discussed in Section 3.3. From the rest of the snapshots in Figure 7 (b), it is clear that the area of influence of the tsunami waves is always confined within the reach of atmospheric pressure waves, because the leading wave is the fastest-travelling one and is locked to the atmospheric pressure wave. At 6 h, the leading and trailing tsunami waves are easily distinguished, since the separation between them is approximately 2,350 km, which is more than twice the wavelength. The higher wave heights continue to propagate towards America, where the highest waves from among the far-field locations were recorded. Interestingly, some waves can be distinguished in the Caribbean Sea in the last panel. These have been generated by the atmospheric pressure wave, after it has propagated over Central America, which acts as a barrier for the waves that were produced on the Pacific Ocean.

Figure 8 illustrates the comparisons of the corrected DART data (black line) and simulated numerical results (red line), arranged from the near-field to the far-field stations. The locations of these stations can be found in Figure 6 (a). The gray arrow marks the arrival of the leading locked wave. The blue dash vertical line indicates the expected arrival time of the trailing free wave, which is calculated as the distance from Tonga to the station over the estimated long wave speed of 756 km/h, calculated for a constant average depth of 4,500 m, representative of the Pacific Ocean. We note the arrival times of the locked and free waves in the near-field (e.g., DART G) are almost simultaneous, because the speed of propagation of the atmospheric pressure wave is small initially and increases to a constant value in the far-field (see Figure 4). The separation (i.e., difference in arrival times) between the leading locked wave and the free wave increases with the distance to the DART station (Liu and Higuera, 2022).

Generally, the simulation results match reasonably well with the corrected DART measurements, both in terms of the arrival time and amplitude of the leading wave. Yet, some differences are clearly observed. In the near-field, at DART G, the tsunami wave has a very deep trough compared to the crest, matching the initial shape of the shock wave (see Figures 3 and 5). After the leading wave, the trailing waves contain higher frequency waves, which are not reproduced by the numerical model. The same feature can be observed at other stations in the near-field (e.g., F, 51425, K, D, etc.). These short waves may have been produced by other tsunami generation mechanisms aside from the atmospheric pressure wave (e.g., the collapse of the volcanic caldera), which have not been considered in this work. This hypothesis seems to be confirmed by the fact that high frequency waves are captured to some extent in Lynett et al.'s (2022) simulations, which accounts for the local wave generation mechanisms. This last point will be further discussed in appendix B. At DART A, which is approximately 2,500 km from the source, the leading wave is well captured in terms of shape and phase, but significant differences are noticeable in the trailing free waves. The general characteristics of the numerical results are in accordance with the analytical solutions of Liu and Higuera (2022), namely, the leading locked wave is led by a crest and the trailing free wave by a trough. The measured trailing wave does not exhibit high-frequency wave components at this particular location, suggesting that the tsunami waves generated by the additional mechanisms may not propagate in all directions equally, and are perhaps influenced by blocking effects or diffraction due to islands, or by preferential propagation directions due to the bathymetry. At this and many other locations the wave amplitudes are smaller than 5 cm in height, which highlights the high sensitivity of the DART system.

In the mid-range-field (Figure 8 (b), 3,000~6,000 km), the simulated leading wave amplitudes and arrival times generally match well with the measurements. Major differences can be found at DART 51407, in which large high-frequency waves are observed in the field but absent in the numerical results.

The analysis of the far-field has been split into three areas. In the Northwest Pacific Ocean region (Figure 8c), the numerical results match especially well with the measurements for the leading waves, and fail to capture the high-frequency oscillations, which in this case start slightly ahead of the estimated arrival time (blue dashed line). These are most likely associated with the additional free waves generated by the interaction of locked waves with changing bathymetry. In the Northeast Pacific Ocean (Figure 8d), the situation is almost identical, but the trailing waves are larger than those in the Northwest Pacific Ocean region. In the Southeast Pacific (i.e., South America, Figure 7e), the separation of the leading locked waves and trailing free waves is more obvious than in other regions, because the propagation distances are longer (Liu and Higuera, 2022). Besides, the amplitude of the trailing waves is significantly larger than that of the leading waves. The main reason is that the DART locations are almost 10,000 km away from Tonga, and the atmospheric pressure waves have decayed significantly, producing a small leading locked wave. Moreover, the trailing free waves are largely amplified when the atmospheric pressure travels over Tonga's Trench (which has a maximum depth of approximately 10,000 m), approaching Proudman resonance conditions (Proudman, 1929), as will be explained in more detail in a later section.

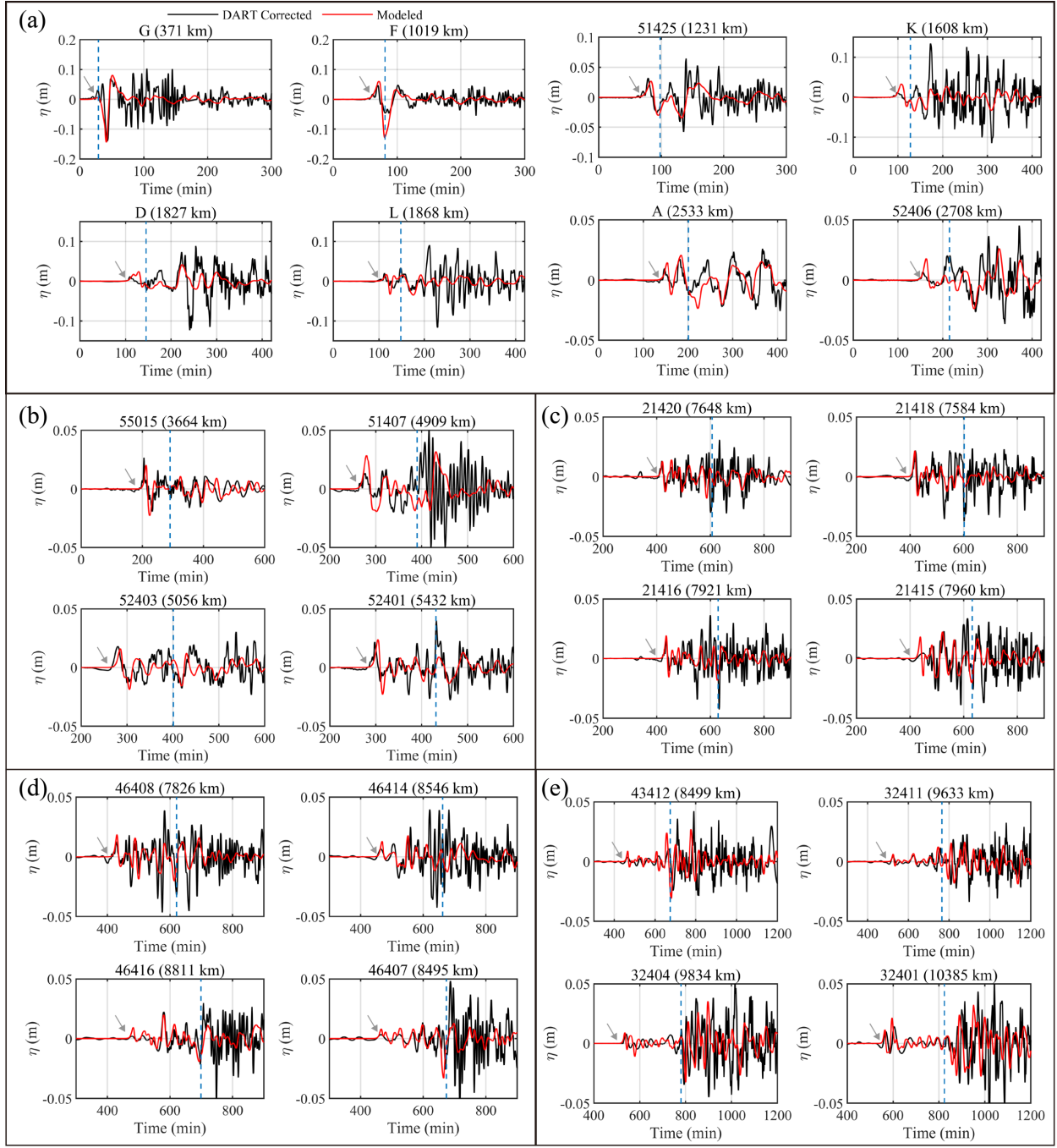

**Figure 8** Comparisons of the measured and the numerically simulated time series of free surface elevations. The black line represents the corrected DART data and the red line denotes the numerical results. The arrival time of the leading wave is marked by the grey arrow and the blue vertical dash line indicates the expected arrival time of the trailing waves. Panel a) contains the results in the near-field (< 3,000 km); b) in the mid-range field (3,000~6,000 km); c), d) and e) in the far-field, for the northwest, northeast and southeast Pacific Ocean, respectively.

To quantify the accuracy of the numerical results for the leading tsunami waves, we compare the arrival time, wave crest and wave trough elevations of the tsunami with the corrected DART measurements in Figure 9. The perfect fit line and the ±20% error lines are shown. The distance between the DART station and the source of the explosion is color-coded according to the color scale shown on the first panel. The results indicate that the numerical model produced accurate solutions for the arrival time, with a coefficient of determination of $R^2 = 0.9994$. This result is not surprising, since the leading wave is "locked" to the atmospheric pressure wave, for which the propagation speed has been derived from observations (see Figure 4). The amplitudes of the leading wave crest and trough at most DART buoys are less than 0.04 m, and they decrease from the near-field to the far-field. The $R^2$ values are 0.75 and 0.78 for the wave crest and trough amplitudes, respectively, when compared with the corrected DART data, indicating that the numerical results are acceptable in terms of the correlation. Comparisons with the uncorrected data yield a slightly higher correlation in terms of $R^2$ (8% and 3% higher), but also significantly larger root mean square errors (RMSE, 98% and 47% higher than for the corrected results, respectively), indicating that the numerical simulation and the simple correction method proposed in Liu and Higuera (2022) provide reasonable results.

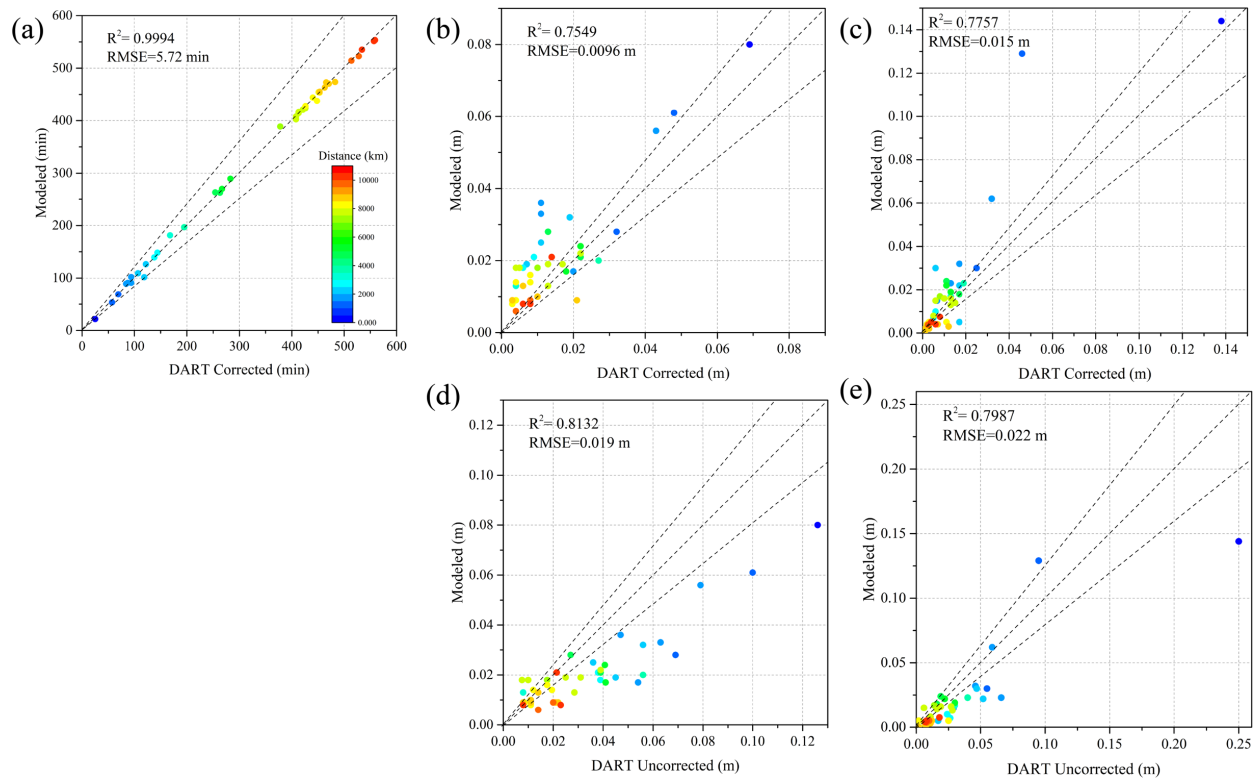

**Figure 9.** Scatter plots showing the model-data comparisons of the arrival time (a), wave crest amplitude (b, d), and wave trough amplitude (c, e) of the leading tsunami waves. Top panels are the comparisons against corrected DART data and bottom panels against uncorrected DART data.

3.2 Wavelet analysis

Using the Morlet (Gaussian-windowed sinusoid) Continuous Wavelet Transformation (CWT) analysis, the time evolution of the dominant wave components of the free surface elevation at selected DART stations are presented in Figure 10. The stations (a-f), are arranged based on increasing distance to Tonga, from top to bottom. The wavelet results based on the corrected DART data are shown in the left column and the results based on the numerical simulations are given in the right column. The arrival time of the leading wave is marked with a white dashed line, while the white continuous line represents the estimated arrival time of the trailing wave. The arrival time of the trailing wave at DART buoy (G) is not shown because it is almost concurrent to the leading wave, as this location is very close to the volcano. At the DART buoy (G), the dominant wave periods range from 10 to 64 minutes initially. Additional components with periods of 5 minutes and shorter appear later, 1 hour from the explosion, as the energy contained in higher period waves reduces progressively, almost disappearing 2 hours after the event. From this point on, the "ringing" effect for high-frequency waves becomes intermittent and continues for several hours. The numerical results are only able to capture the low frequency waves, with a slightly smaller block (15 – 40 minutes) centered around the 28 minutes period, since additional wave generation mechanisms have not been considered in the present model. This situation can be observed in all subsequent numerical modelling results, which lack short period waves below ~15 minutes. Moreover, the numerical results at DART G show an additional long period component at 70 minutes, with higher energy content, when compared to the field data. A common factor between the DART data and the numerical model data at DART G (the closest to the volcano) is that energy exists before the arrival of the leading wave. However, these points appear in both panels, and lie outside the cone of influence of the CWT (yellow curved dashed line). For those reasons, they should be considered as an energy "leakage" instead of actual long wave conditions at the site prior to the explosion.

Based on the results, the numerical simulation based on the simple atmospheric model can successfully reproduce in a qualitative way the longer period waves (40 minutes at DART 51425, 56 minutes at DART 46407 and 40-69 minutes at DART 32401), leading or trailing. This conclusion is consistent with the observations already made for Figure 8. Generally, the agreement is higher in terms of the starting times (relative location within the figure) than in the magnitude (color scale). However, in some instances, such as close to Tonga (DART G, 51425) and in the Northwest Pacific Ocean (DART 21420), the differences are significant, especially when comparing the components with a higher energy content. In fact, the field observations of DART 51425 highlight the extreme complexity of this event in the near field, since model fails to capture most of the higher energy components at this location, especially after the trailing waves arrive. This is not a surprise, since as already mentioned, higher frequency waves are not included in the source model. As we move farther away from the source, the leading and trailing waves are further separated in time, and it is easier to observe the differences between them. The amplitude of the leading wave decreases with distance, which is reflected in the CWT results. Moreover, the leading wave tends to produce long waves only, and little or no shorter wave components. In some instances (e.g., DART 21420) lower period components arrive before the trailing wave arrival time, which means that these must have been produced locally or closer to the DART buoy location rather than at the volcano, due to their slower celerity. These are likely to have have been produced by wave scattering (i.e., free wave creation) as the locked wave propagates over the changing bathymetry, by the mechanisms explained in Vennell (2007). Given that the model solution does not capture these short-wave components (e.g., compare both panels in row d), the main hypothesis is that they may require a finer mesh to be reproduced. However, the finer mesh tested led to numerical instabilities, so this point could not be verified. On top of this, components with shorter wavelengths are more susceptible to dispersive effects. Therefore, LSWE may not suffice to represent them accurately, in which case more sophisticated types of equations, as for example Boussinesq equations, would be required. An analysis on the dispersion effects for the present model will be presented in appendix A. Finally, after the trailing wave arrival time, long and short waves seem to be equally important. This is because the free wave produced by the atmospheric pressure travels at the same celerity as long waves produced by additional mechanisms. Generally, the longer period components arrive slightly earlier. However, this perception may be slightly exacerbated by the characteristics inherent to the CWT, which spreads energy over a longer time for longer periods, due to the longer length of the wavelet used in the analysis of longer periods.

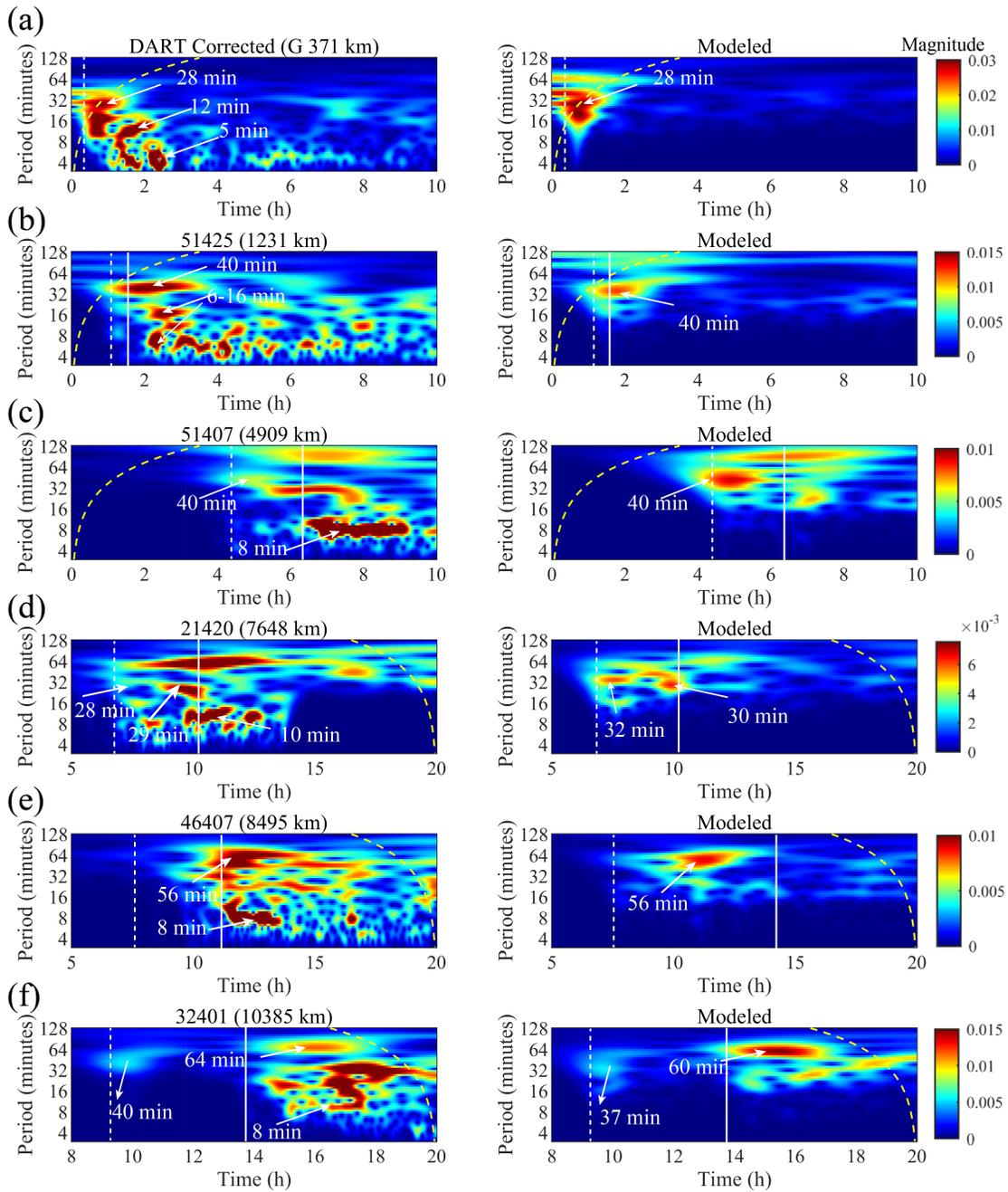

**Figure 9** Wavelet analysis of the time series of surface elevation using the corrected DART measurements (left column) and the numerical results (right column), arranged from the near field to the far-field. High energy components highlighted in red and lack of energy in blue.

3.3 Tsunami wave amplitude distribution in the Pacific Ocean

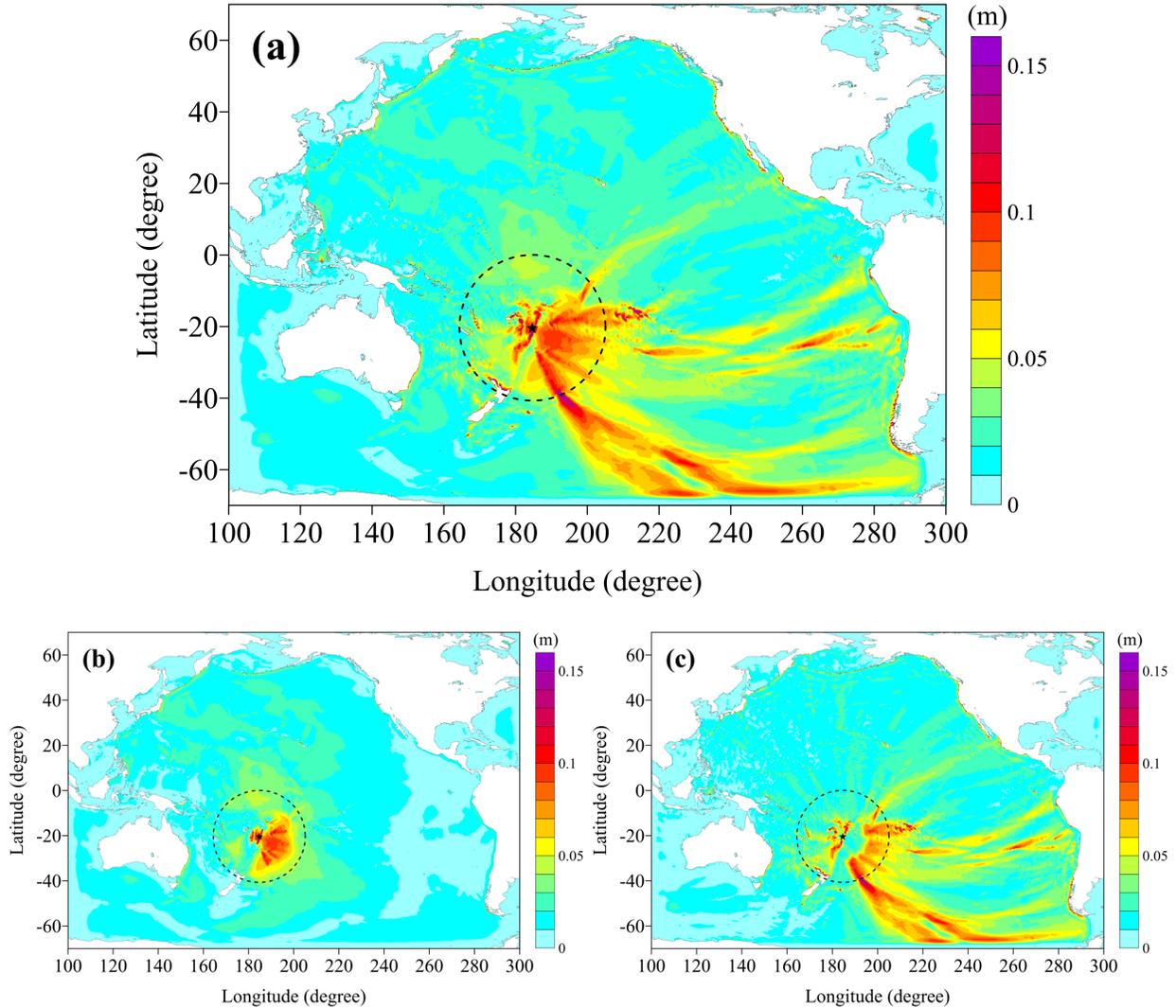

**Figure 11 Distribution of the maximum tsunami amplitudes on the simulation. (a) Overall maximum amplitudes; (b) maximum amplitudes of the leading wave; (c) maximum amplitudes of the trailing wave. Tonga's volcano denoted by a black star. The dashed line marks the location after which the leading and trailing waves are fully separated.**

The spatial distribution of the maximum (positive) tsunami amplitude in the Pacific Ocean numerical simulation is presented in Figure 11. Panel (a) shows the global maximum throughout the numerical simulation, while panels (b) and (c) present the maximum values associated with the leading and trailing waves, respectively. Therefore, panel (a) can be viewed as the maximum among panels (b) and (c). To separate the influence of the leading and trailing waves, panel (b) records the maxima at locations, which are within the influence of the pressure disturbance (r >

$X_l(t)$ – L/2 , with L being the wavelength, 900 km), whereas panel (c) records the maxima elsewhere. Both the leading and trailing waves are produced at the same time and location (the explosion at the volcano) but propagate at different celerities and the distance between them increases progressively. Thus, there is an area near Tonga in which they overlap, which is bounded by the dashed line circle. At any location outside the circle, which has a radius of approximately 2,200 km (based on the pressure wave celerity, 319 m/s, free wave celerity, 198 m/s, and wavelength, 900 km), the leading and trailing waves can be perfectly distinguished.

The wave amplitudes linked to the leading wave, shown in panel (b), are significantly larger in the near-field, where the atmospheric pressure is higher (see Figure 3). The maximum amplitude measured is approximately 10 cm. The area with the highest amplitudes is linked to the Tonga trench, where the depth is in the excess of 10,000 m (see the dark shade of blue line east of Tonga in Figure 1). Therefore, the local wave celerity of the free waves over the trench is roughly 313 m/s, while the speed of atmospheric pressure waves is about 319 m/s, producing a Froude number very close to 1 (the critical condition), which is commonly associated with the Proudman resonance condition (Proudman, 1929). This effectively translates in an increase of the rate at which the atmospheric pressure disturbance transfers energy to the sea, which in turn produces a higher response of the free surface (i.e., larger waves). In addition, locked waves propagating over steep bathymetry features (e.g., ridges and trenches) are known to produce scattering of free waves (Vennell, 2007). These two processes can explain the fact that the wave height is significantly larger in the vicinity of Tonga's trench in panel (b), and especially on the east side, which is "downstream".

Regarding the tsunami wave amplitude associated with the trailing waves, shown in panel (c), it is clear that the amplification effect over the Tonga Trench plays a significant role in the large amplitudes observed in the Southeast Pacific as well. The amplitudes of the trailing waves in the North-West Pacific are negligible (generally smaller than that of the leading wave). Since the pressure forcing function is perfectly uniform in space and extends as a perfect circle with its center at Tonga, this indicates that the trailing wave produced initially by the pressure model is not likely to have produced the large waves in the South-East Pacific. Instead, it is highly likely that those have been produced by the free scattered waves generated as the pressure wave propagates over Tonga Trench instead (Vennell, 2007), as just mentioned. These free waves travel at the long wave celerity, which is slower than the pressure wave celerity and progressively separate from it.

Moreover, because they have been generated some time after the explosion and some distance away from Tonga, their expected arrival time is still prior to the expected (original) trailing wave arrival. Therefore, wave scattering is most likely the reason why in Figure 8 some of the stations present significant waves arriving after the leading wave, but before the expected arrival time of the trailing wave. Moreover, it can also be observed that in the Northwest Pacific, which is free from the effects of Tonga Trench, the highest amplitude overall is produced by the leading wave.

In addition to the larger wave heights in the Southeast Pacific, panel (a) presents interesting finger-like features in which the maximum wave amplitude is larger. Upon closer inspection, these may have been produced by wave concentration due to the bathymetry features. For example, the south-most feature appears to correlate with the location of the Louiseville Ridge, east of New Zealand. The fingers channeling wave energy east are likely enhanced by the Manihiki Plateau (2,500 m water depth in an area that is almost 5,000 m deep), the Tuamotu Archipelago, before following the Sala y Gomez and Nazca ridges towards the coasts of Peru, where two people drowned and an oil-spill occurred due to the 2 m high tsunami waves.

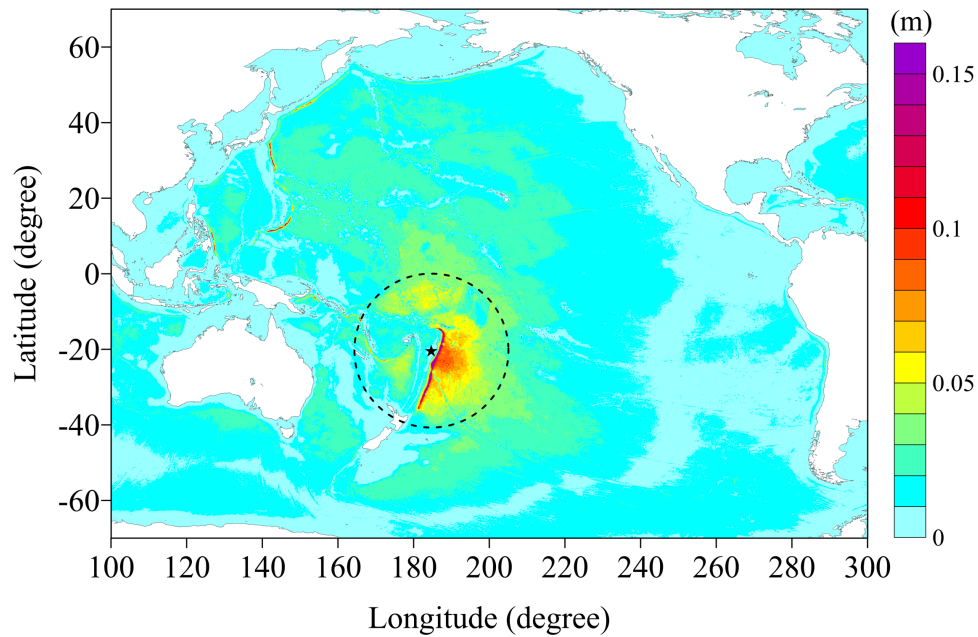

**Figure 12 Distribution of the maximum tsunami amplitudes of the leading wave based on the formulation in Liu and Higuera (2022). See Figure 11 caption for further context.**

It is well known that under simple conditions (i.e., constant water depth, shallow water wave regime) the ocean locked wave has the same shape as the pressure wave, and its amplitude is proportional to the pressure times the amplification factor $\frac{1}{Fr^2-1}$ (Vennell, 2007; Liu and Higuera, 2022). Applying this expression considering the maximum positive amplitude of the pressure model as a function of the distance to Tonga (Equation 5) and the local depth at each point in space to calculate the amplification factor, a simple estimator of the leading wave positive amplitude can be constructed. The theoretical distribution of the maximum leading wave is shown in Figure 12. We remark that this is a crude approximation, since it assumes an immediate response of the water and there are additional physical processes such as wave scattering, shoaling, refraction, diffraction, etc., which are not captured. The processes, however, are correctly captured by the numerical simulation. Consequently, it interesting to compare Figure 12 with panel (b) in Figure 11 to evaluate the main differences.

To start with, a large amplification factor of approximately 25 is obtained over Tonga's Trench, leading to the largest theoretical amplitudes of the locked wave, exceeding 1 m. Other trenches, like Mariana, Japan and Kuril, located in the Pacific Nortwest, also show potential large wave heights due to the same mechanisms. However, pressure has decayed significantly by the time it reaches those locations, leading to lower wave heights overall. Despite, the shortcomings of the simplified analysis, the comparison with panel (b) in Figure 11 presents similarities. The main differences occur inside the dashed line circle, where the locked and free waves overlap, and where wave scattering is significant. Because of this, the wave heights in the vicinity of Tonga (black star) are much larger in the numerical model solution. At the east side of Tonga's Trench, the degree of similarity increases, although the numerical model solutions present a larger wave over a wider geographical area, for the aforementioned reasons. Outside the circle, where the influence of the leading wave can be isolated from the effects of the trailing waves, the degree of similarity is outstanding, matching not only in terms of the general patterns, but also in terms of the final amplitude. Therefore, it can be concluded that the analytical model of Liu and Higuera (2022) is applicable to real conditions, especially in areas with small bathymetry gradients.

## 4 Conclusions

In this study, we first constructed an N-wave function to describe the atmospheric pressure characteristics induced by the eruption of the 2022 Hunga Tonga-Hunga Ha'apai Volcano. The empirical model was calibrated with field measurements. The atmospheric pressure crest amplitude increases from the source in the near field, then decreases slowly moving aways from the source. On the other hand, the trough amplitude of the atmospheric pressure decreases quickly from the source in the near-field, and then slowly in the far-field. The propagation speed of the atmospheric pressure increases in the near-field and becomes a constant speed of 319 m/s at about 1,000 km from the volcano. The method suggested by Liu and Higuera (2022) is adopted to correct the DART measurements to exclude the atmospheric pressure influence in the signal for the free surface elevation.

Based on the LSWE model, the simulated tsunami waves results match reasonably well with the corrected DART measurements in terms of the tsunami arrival time and tsunami amplitude from the near-field to the far-field, especially for the leading wave and most parts of the trailing wave. However, the short-wave components of the trailing waves, which could possibly be generated by another generation mechanism, cannot be captured by the present numerical model. The wavelet analysis for the corrected DART data and the numerical results supports this point. We successfully separate the tsunami leading wave and trailing wave numerically and theoretically. The larger atmospheric pressure amplitude and Tonga Trench increase the tsunami leading wave in the near field. The amplification effect induced by Tonga Trench plays an important role in the large amplitudes in the Southeast Pacific. We also investigated the importance of the frequency dispersion effect. The results show that the dispersion effect is minor for the 2022 Tonga event, due to its long wavelength of 900 km. We further compared our results with Lynett et al. (2022)'s model, as we have applied similar method. Both models capture the pressure forcing mechanism, including the trailing waves in the far-field. This indicates that a significant part of the trailing waves is induced by the atmospheric shock wave. In conclusion, the method suggested in the study could explain the tsunami generation and propagation generated by the 2022 Tonga volcano eruption in the Pacific Ocean. Nevertheless, the other possible mechanisms should be investigated to explain the trailing waves with shorter periods.

**Acknowledgments**

P. L.-F. Liu would like to acknowledge the National University of Singapore research grant (NRF2018NRF-NSFC003ES-002). Z. R. Ren would acknowledge the National Natural Science Foundation of China (Grant Nos. 12002099). This study was supported by the Yushan Program, Ministry of Education in Taiwan.

**Data availability statement**

The bathymetry data used in this work are derived from the ETOPO1 global relief model (https://maps.ngdc.noaa.gov/viewers/grid-extract/index.html). The atmospheric pressure measurement is obtained from Fiji Meteorological Service (https://www.met.gov.fj/), Weather Underground (https://www.wunderground.com/), NIWA (https://niwa.co.nz/), Japan Soratena (https://global.weathernews.com/news/16551/), Meteorological Directorate of Chile (https://climatologia.meteochile.gob.cl/), Central Weather Bureau of Taiwan (https://www.cwb.gov.tw/), and National Climatic Data Center of USA (https://mesonet.agron.iastate.edu/request/asos/1min.phtml#). The DART buoy data is provided by the National Oceanic and Atmospheric Administration of USA (https://nctr.pmel.noaa.gov/Dart/).

Appendix A. Frequency dispersion effects

To investigate the significance of frequency dispersion on the tsunami wave generation and propagation during the Tonga event, an additional numerical simulation, using the linear Boussinesq equations has been performed.

Before presenting the numerical results, Glimsdal et al. (2013) introduced the parameter $\tau$, known as "dispersion time", to assess the potential effect that dispersion has on wave propagation:

$$\tau = \Delta c \cdot t \cdot \frac{1}{L} \approx \frac{6c_0 d^2}{L^2} \cdot t \cdot \frac{1}{L} = \frac{6d^2 D}{L^3} = \frac{6dt}{gT^3}, \tag{13}$$

where $d$ is water depth, $L$ is wavelength, $c_0$ is water wave celerity, $T$ is wave period and $D$ is propagation distance. In this study, the characteristic wavelength $L$ is 900,000 m and the average water depth is approximately 4,500 m, which is representative of the Pacific Ocean. The effects of wave frequency dispersion accumulate in time/space. Therefore, according to Glimsdal et al. (2013), if $\tau < 0.01$, the dispersion effects are small, and if $\tau > 0.1$, the dispersion effects become significant. For a propagation distance of 10,000 km, which is the order of magnitude of the width of the Pacific Ocean, $\tau$ is 0.0016. This suggests already that the frequency dispersion effects for the Tonga event are expected to be negligible.

The time series of the surface elevation at two DART buoys in the far-field comparing the dispersive and non-dispersive simulations are plotted in Figure 13. The results show minor differences only, with deviations being less than 6%. This indicates that the dispersion effect can be safely ignored when simulating the wave driven by the moving pressure in the 2022 Tonga tsunami event, since the pressure shock wave is a very long wave. This conclusion, however, does not hold when reproducing the other wave generation mechanisms. The additional mechanisms produce significantly shorter waves, as has been shown in Lynett et al. (2022). Therefore, shorter waves will be more affected because they travel a longer distance relative to their wavelength, as will be further discussed in the next section.

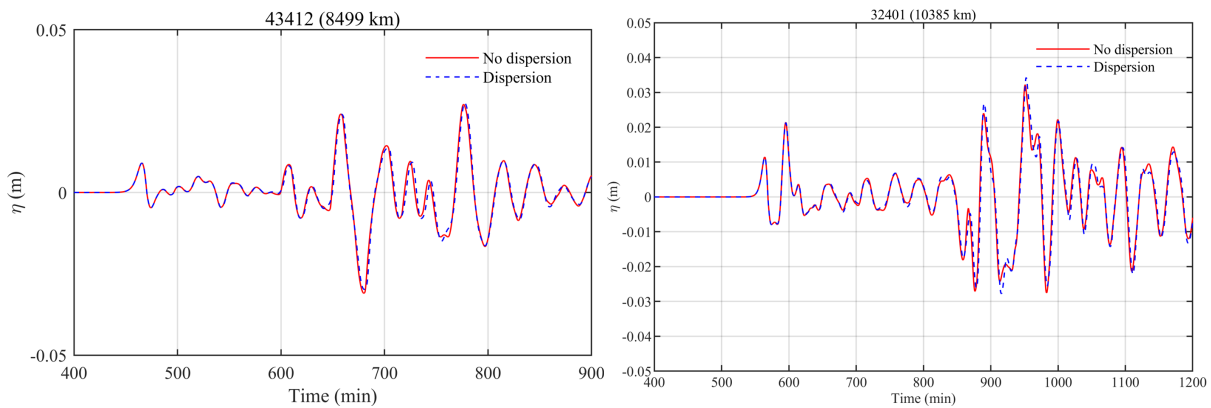

Figure 13 Time series of the surface elevation without and with dispersion effect

Appendix B.   Comparisons with Lynett et al.'s model

Lynett et al. (2022) have presented an alternative model to describe the atmospheric pressure and numerical simulations of the induced tsunami. The atmospheric pressure wave is formulated by defining the crest and trough independently to produce the distinctive N-wave shape. The crest and trough present the same basic shape (exponential curves decaying with the square of the phase), but each one has its own amplitude, celerity and wavelength (or wave period). These three are functions of time and that have been found using the available atmospheric pressure data. Lynett et al. (2022) adopted pressure data obtained from 134 weather stations with a sampling frequency of 6-10 minutes, located all around the globe. Thus, the measurements from some of those stations may be affected by the surrounding topographical features, unlike those carefully selected in this work. Lynett et al's model offers a high flexibility, with 6 main parameters, and can account for the changes in wavelength of the pressure wave. In contrast, the model proposed in this paper (Equations 1-6) offers only 3 parameters (the amplitudes of crest and trough and the joint celerity), considering a constant wavelength.

Apart from the differences in the empirical atmospheric models, there are significant differences in the simulation procedure, as follows. On the one hand, while Lynett et al.'s model adopted a Cartesian coordinate system based on the azimuthally equidistant projection, the present model uses spherical coordinates, which is a more physical representation of reality. In addition, we have accounted for Coriolis force. On the other hand, Lynett et al.'s model considered additional wave generation mechanisms (e.g., caldera collapse) on top of the atmospheric pressure wave, which introduce shorter waves in the system, whereas the present model only considers the latter. As a result, Lynett et al.'s model is based on the Boussinesq equations, which account for the frequency dispersion effect, while the present model adopts a (non-dispersive) LSWE model.

In order to allow a direct comparison between the performance of both models, Lynett et al.'s results have been analyzed and presented in Figure 14, which can be directly compared with Figure 9. In both cases the arrival time data is virtually the same, and very close to a perfect fit. Regarding crest and trough amplitudes, the $R^2$ is very similar or slightly better in Lynett et al.'s (2022) case, while the root mean square error (RMSE) metrics are higher by approximately 20% and 60% than in our case. However, Lynett et al.'s (2022) data includes only 18 points, so any outlier will have more relative weight, whereas in this work we have compared data from 41 DART stations.

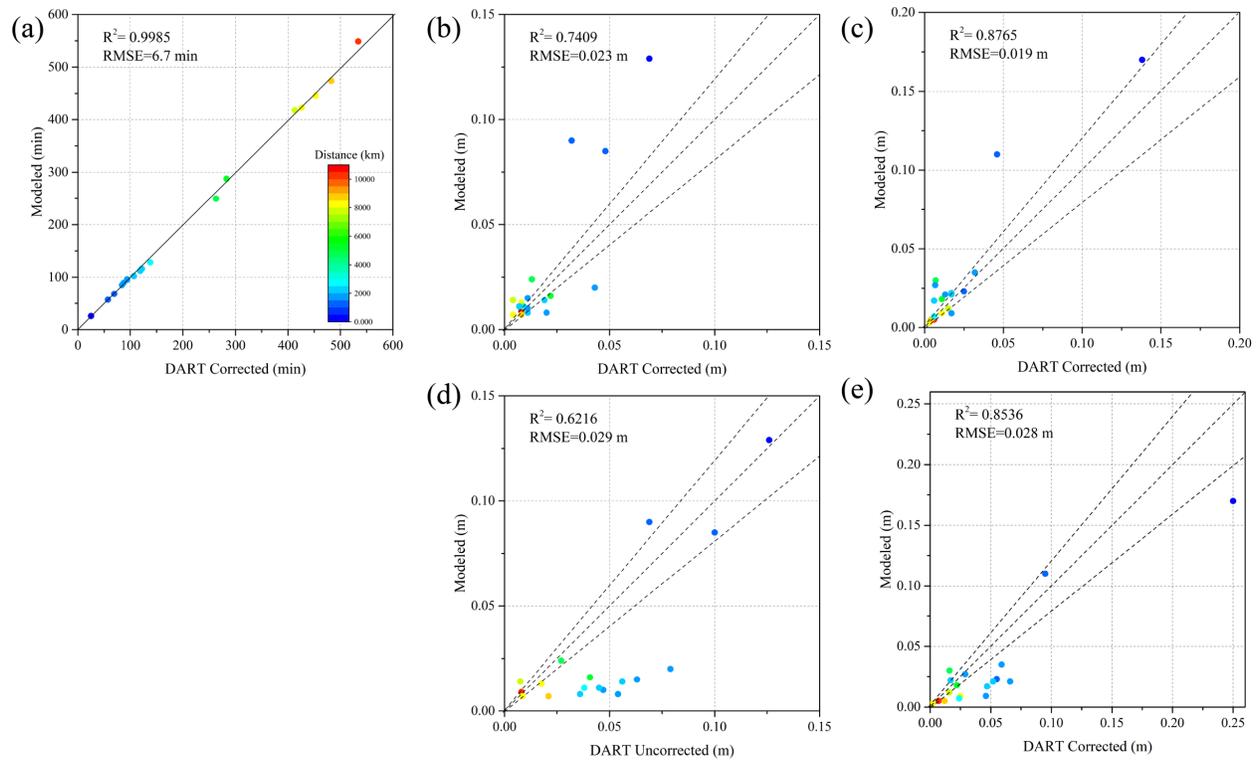

**Figure 14. Scatter plots showing the comparison of results of Lynett et al. (2022) in terms of arrival time (a), wave crest amplitude (b,d) and wave trough amplitude (c,e) of the leading tsunami. Top panels show comparisons against corrected DART data and bottom panels against uncorrected DART data. This figure can be directly compared with Figure 9.**

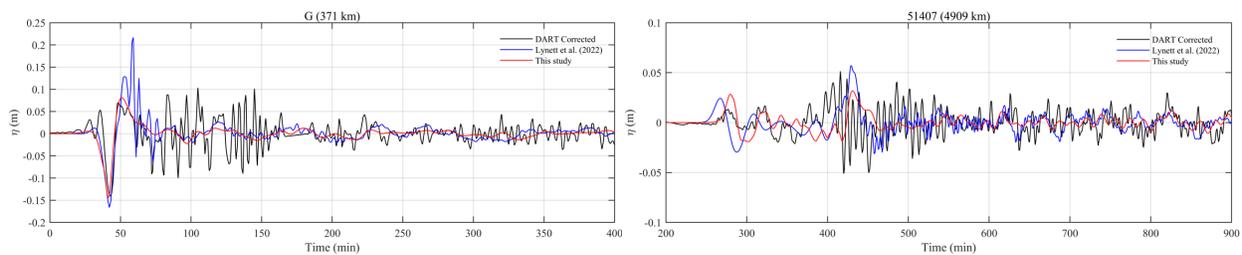

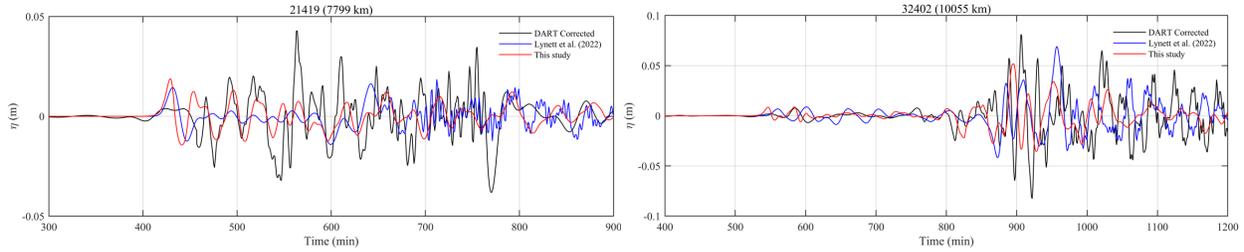

**Figure 15.** Comparison of time series of free surface elevation between present results (red line), Lynett et al.'s (2022) results (blue line) and the corrected DART data (black line).

We have also compared the time series of free surface elevation between the present numerical simulation results (red line) and Lynett et al.'s (2022) results (blue line), as shown in Figure 15. In the near field (DART G), both solutions are very similar in terms of the long-wave components. However, Lynett et al.'s solutions present noticeable shorter wave components arriving 50 minutes after the volcanic explosion, which are likely produced by the additional mechanisms included in their model. Nevertheless, these shorter waves do not match with the arrival time observed in the field, which happens approximately 25 minutes later. In the mid-field, the amplitudes of the crest and trough of the leading wave for both models in DART 51407 are very similar, and slightly higher than the DART corrected data. However, the arrival time from Lynett et al.'s solution is well ahead of the observed arrival time, whereas the model presented in this paper captures the arrival time almost perfectly. The long-wave component (i.e., trend) of the trailing waves is also similar in both cases. However, the present model does not capture the high-frequency waves (as expected) and Lynett et al.'s results show a later arrival and a smaller amplitude than in the field data. Also in the mid-range, DART 21419, neither model presents accurate results, especially after the first waves, reflecting the complexity of the real world problem.

Finally, in the far-field (DART 32402) both models produce similar results and are able to model the arrival time of the trailing waves accurately, capturing their large amplitudes to a great extent. This suggests that the trailing free waves are produced at the different stages and locations during the simulation, when the locked waves travel over the changing bathymetry, as has already been discussed. The additional wave generation mechanisms (e.g., caldera collapse) have some impact on the solutions, especially in the short-wave range in the near-field, but are likely to have a second order effect farther away, since the trailing waves in the far-field have been found to be very similar in both cases.